\documentclass[pop,numberedappendix,iop]{aeb_emulateapj_2015}
\usepackage{amsmath,cases}
\usepackage{color}
\usepackage{longtable}
\usepackage{ulem,array}
\usepackage{grffile}
\usepackage{cleveref}
\usepackage{multirow}

\usepackage[mathscr]{euscript}

\definecolor{epcol}{rgb}{0.398, 0.0, 0.797}

\begin{document}

\author
{
Mohamad Shalaby\altaffilmark{1,2,3}, 
Avery E. Broderick\altaffilmark{1,2}, 
Philip Chang\altaffilmark{4},
\\
Christoph Pfrommer\altaffilmark{5,6}, 
Astrid Lamberts\altaffilmark{7}
and
Ewald Puchwein\altaffilmark{8}}

\altaffiltext{1}{Department of Physics and Astronomy, University of Waterloo, 200 University Avenue West, Waterloo, ON, N2L 3G1, Canada}
\altaffiltext{2}{Perimeter Institute for Theoretical Physics, 31 Caroline Street North, Waterloo, ON, N2L 2Y5, Canada}
\altaffiltext{3}{Department of Physics, Faculty of Science, Cairo University, Giza 12613, Egypt}
\altaffiltext{4}{Department of Physics, University of Wisconsin-Milwaukee, 1900 E. Kenwood Boulevard, Milwaukee, WI 53211, USA}
\altaffiltext{5}{Heidelberg Institute for Theoretical Studies, Schloss-Wolfsbrunnenweg 35, D-69118 Heidelberg, Germany}
\altaffiltext{6}{Leibniz-Institut f{\"u}r Astrophysik Potsdam (AIP), An der Sternwarte 16, D-14482 Potsdam, Germany}
\altaffiltext{7}{Theoretical Astrophysics, California Institute of Technology, Pasadena, CA 91125, USA}
\altaffiltext{8}{Institute of Astronomy and Kavli Institute for Cosmology, University of Cambridge, Madingley Road, Cambridge, CB3 0HA, UK}

\email{mshalaby@live.ca}

\title{Importance of resolving the spectral support of beam-plasma instabilities in simulations}

\shorttitle{Importance of resolving the spectral support of beam-plasma instabilities in simulations}
\shortauthors{Shalaby et al.}

\begin{abstract}

Many astrophysical plasmas are prone to beam-plasma instabilities.
For relativistic and dilute beams, the {\it spectral} support of the beam-plasma instabilities is  narrow, i.e., the linearly unstable modes that grow with rates comparable to the maximum growth rate occupy a narrow range of wave numbers.
This places stringent requirements on the box-sizes
when simulating the evolution of the instabilities.
We identify the implied lower limits on the box size imposed by the longitudinal beam plasma instability, i.e., typically the most stringent condition required to correctly capture the linear evolution of the instabilities in multidimensional simulations.
We find that sizes many orders of magnitude larger than the resonant wavelength are typically required.
Using one-dimensional particle-in-cell simulations, we show that the failure to sufficiently resolve the spectral support of the longitudinal instability yields slower growth and lower levels of saturation,
potentially leading to erroneous physical conclusion.

\end{abstract}

\section{Introduction}

Plasmas are ubiquitous in astronomical environments, from the intergalactic medium to gamma-ray bursts.  Often, these are subject to virulent plasma instabilities that couple the microscopic motions of the constituent particles and the macroscopic, collective motions of the fluid.
These are believed to be critical to mediating collisionless shocks, growing magnetic fields, coupling different particle species, and accelerating a fraction of these particles to high energies.

Beam-plasma instabilities, in particular, are a common feature in astrophysical contexts. They typically occur when relativistic, dilute beams propagate through background plasmas.
Examples include
relativistic jet propagation~\citep{Jet-sims2,Jet-sims1},
gamma-ray bursts~\citep{GRB+BP-instability1,GRB+BP-instability2},
collisionless Shocks~\citep{shock1,shock3,shock2}, 
black hole accretion flows~\citep{accretion1},
and pair beams, induced by TeV-Blazars~\citep{blazarI,Schlickeiser+12,Schlickeiser+13,nonlinear-paper,linear-paper}, propagating through the intergalactic medium.

Linear perturbative analyses for typical astrophysical plasmas show that a subset of linear plasma wave modes are unstable, leading to the exponential growth of their amplitudes.
However, analytical solutions for linear dispersion relations can be found only in very simple cases
and depend on the particulars of the equilibrium momentum distribution of plasmas~\citep{Bret2010}.

These instabilities back-react on the particle distribution that originally drove them in the linear regime and nonlinearly saturate.  
Analytical analysis of this nonlinear saturation rely on making appropriate choices of the physics to include, as well as on making certain approximations and simplifications of that physics. 
As a result, these analyses can sometimes give contradictory conclusions, \citep[e.g.][]{Miniati-Elyiv-2013,nonlinear-paper}.  Numerical simulations offer a path to clarify the physics of nonlinear saturation, but present their own caveats and subtleties.

The particle-in-cell (PIC) method \citep{Hockney-Eastwood,Birdsall-Langdon} is a powerful and commonly used numerical tool to study the kinetic evolution of plasmas.
The general idea of this algorithm is straightforward: it follows the trajectory of particles with $N$-body methods while solving Maxwell's equation on an Eulerian grid.
Thus, the method simultaneously takes advantage of the nearly uniform spatial extent of most plasmas (i.e., the large separation in scales between the plasma skin depth and the typical length scales of astrophysical systems) and addresses the sparse local momentum distributions that are usually relevant.  The two partial descriptions of the system are coupled via interpolation, both to deposit charges and currents and to obtain the forces on the particles.  While most of our remarks are true for simulation methods in general, we will explicitly demonstrate them for PIC simulations in one dimension.

The nature of the PIC algorithm, which combines a discretization of momentum and physical space, evokes three notions of  ``resolution'' that are important to accurately model the evolution.
First is spatial resolution, e.g., the cell size on the Eulerian grid, $\Delta x$.
Insufficient spatial resolution necessarily leads to artificial numerical damping as the various plasma waves can no longer be resolved.
Second is momentum-space resolution, e.g., the number of particles per cell, $N_{\rm pc}$, in PIC algorithms.
This is conceptually similar to spatial resolution, being essentially part of the generalization of the same notion to the phase space.
Insufficient momentum space resolution leads to the inability to resolve current fluctuations, again leading to unphysical plasma evolutions.
Typically, simulations are tested for convergence by decreasing $\Delta x$ and increasing $N_{\rm pc}$. It was show in~\citet{sharp} that simulations only converge when these are changed concurrently.

A third, often ignored, notion is {\it spectral} resolution, which is necessary to resolve all the relevant modes in the system. 
Here, we show that, within the context of simulating beam-plasma instabilities, this notion is just as important as the preceding two.
Typically, simulations of the instabilities' saturation employ periodic boundary conditions, for which only wave modes that fit fully within the simulation box are modeled.  
That is, a box with length $L$ supports only modes with wave numbers
\begin{equation}
k_j = j \Delta k_{\rm sim}
~~\text{where}~~
\Delta k_{\rm sim} \equiv \frac{2\pi}{L},
\end{equation}
where $j$ is an integer ranging from zero to $N_{\rm c}$.  Thus, the spectral resolution, $\Delta k_{\rm sim}$, is a measure of the finest structures in $k$-space that can be probed by the simulation.  
Anywhere that instabilities have compact support in $k$, e.g., resonant instabilities in cold plasmas, generally, insufficient spectral resolution can lead to their unphysical quenching and sub-optimal growth.

Beam-plasma systems often fall into this pathological category.
That is, only modes with wave numbers within $\Delta k_{\rm f}$ about the fastest growing mode, located at $k_{\rm f}$, grow at rates comparable to the maximum growth rate, $\Gamma_{\rm f}$.  
Therefore, a necessary condition for resolving the instability is
\begin{eqnarray}
\label{eq:spec-res}
\Delta k_{\rm sim} \ll \Delta k_{\rm f }
~~\Rightarrow~~
L \gg \frac{2\pi}{\Delta k_{\rm f}}.
\end{eqnarray}

The severity of this condition depends on the width $\Delta k_{\rm f}$.
For relativistic and dilute beam-plasma systems, the most relevant for many astrophysical plasmas, only a very narrow spectral subset of the unstable linear wave modes grow with rates comparable to the maximum growth rate \citep{Bret2010}.
Hence, this case is a prime example in which the spectral resolution is very important, and a special care must be taken to resolve the narrow spectral support of the important unstable wave modes.

Typically, simulations of beam-plasma instabilities done in literature require that the box size to be larger than the most unstable wavelength, i.e.,
\begin{equation}
 \Delta k_{\rm sim} < k_{\rm f}.
\end{equation}
However, this necessary condition does not guarantee that the simulation can resolve any wave modes within $\Delta k_{\rm f}$, the most important wave modes during the linear evolution.

\setcitestyle{notesep={; }}
 
For relativistic, dilute beam-plasma system, the width of the oblique instability in the cold limit, associated with modes that propagate at an angle to the beam direction, typically  is considerably larger than that for the longitudinal modes\footnote{In the warm beam limit, it is no longer clearly true that the width of the oblique instability is larger than that of the longitudinal instability.  When finite temperature effects are included, the growth maps of the instabilities become much narrower~\citep[see, e.g., the various panels in Figure 17 of ][]{Bret2010}. Therefore, in warm cases, the spectral resolution requirements are likely to impose an even a stronger requirement on the box sizes.}.
However, when considering these multidimensional instabilities, resolving only the oblique modes is known to be insufficient.
In practice, all unstable modes grow simultaneously and impact the subsequent nonlinear evolution. It was shown by \citet[][see especially the discussion in Section V.D]{Bret2010} that,
while the oblique modes dominate initially, upon exiting the linear regime, subdominant modes continue to grow and prove essential to correctly capturing the transition to the fully nonlinear regime.

\setcitestyle{notesep={, }}

We show below that the requirement in Equation~\eqref{eq:spec-res} for longitudinal modes can be very stringent and  imposes a strong restriction on the simulation setup. 
This stringent requirement must continue to be respected in higher-dimensional simulations as well.

The paper is organized as follows:
in Section~\ref{sec:resolution}, we distinguish between different plasma instabilities according to the spectral support of their growth rates, $\Delta k_{\rm f}$, as broad and narrow supported instabilities.
In Section~\ref{sec:simulations}, we show the implications of the narrow spectral support on the growth rates in simulations and quantify the spectral width of unstable longitudinal modes. We then present a number of simulations that show a quantitative agreement between simulation results and such anticipated implications. Concluding remarks are given in Section~\ref{sec:conclusion}.

\section{Spectral support for Plasma instabilities}
\label{sec:resolution}

Here, we distinguish between two different types of instabilities that can occur in astrophysical plasmas: those with broad spectral support and those with narrow spectral support. In wave number space, the distinction is based on the spectral width of modes that grow with rates near the maximum growth rate.
Many different (especially non-relativistic) plasma instabilities exhibit a broad spectral support, but a subclass of important problems are plagued by instabilities with narrow spectral support. 
Assessing when this is likely to be the case is critical to designing appropriate simulations.
Here, we focus on relativistic plasma instabilities: the two-stream instability, which has broad support, and the beam-plasma instability, which has narrow support.

\subsection{Broad spectral support: two-stream instability}

A common example of instabilities with broad spectral support is the two-stream instability, which occurs when two populations of collisionless plasmas of equal density counter-stream through each other with the same speed $v_s$. A subset of longitudinal plasma waves become unstable and grow exponentially with time. This leads to the heating of both streams and the acceleration of a subset of particles.

In the cold limit, the linear dispersion relation of the two-stream system is~\citep[e.g.,][]{2s-disperions} 
\begin{eqnarray}
2 \gamma_s^3 = \frac{ \omega^2_p }{ (\omega + k v_s)^2} + \frac{ \omega^2_p }{(\omega - k v_s)^2},
\label{eq:dis_2stream}
\end{eqnarray}
where $\omega_p = \sqrt{ q^2 n / m \epsilon_0}$ is the plasma frequency, $n$ is the total number density of both streams, $q$ and $m$ are the elementary charge and mass of streaming plasma particles, respectively, and the Lorentz factor of the streams is $\gamma_s = ( 1-(v_s/c)^2)^{-1/2}$.

\begin{figure}
\center
\includegraphics[width=8.6cm]{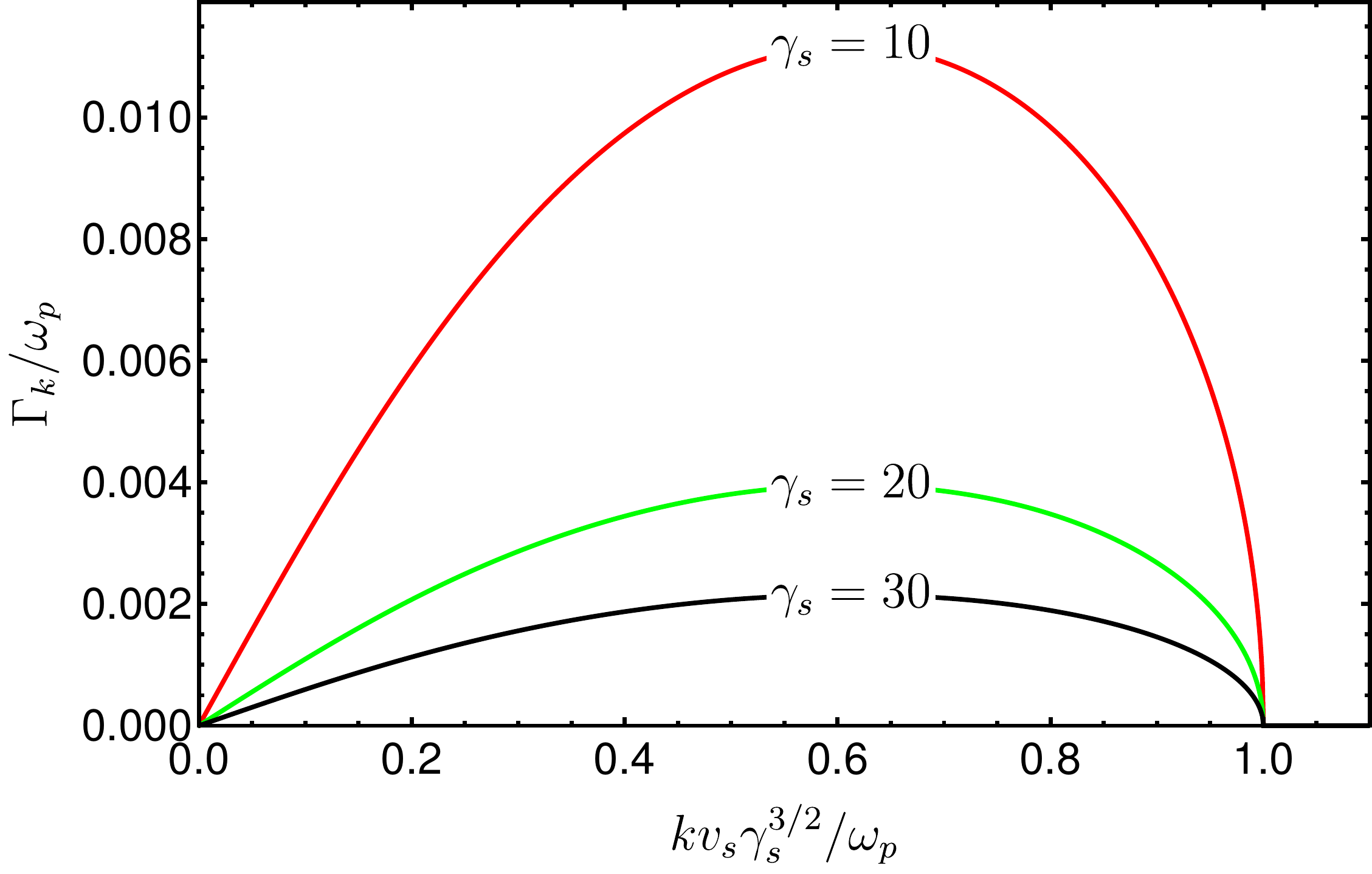}
\caption{
Growth rates $\Gamma_k$ of the unstable wave modes in two-stream system with different center of mass Lorentz factor $\gamma_s$.
Here, $k v_s \gamma_s^{3/2}/\omega_p = \lambda_{\rm min}/\lambda$, where, $\lambda$ is the wavelength corresponding to the wave mode $k$, and  $\lambda_{\rm min}$ is the shortest unstable wavelength.
\label{fig:2stream}
}
\end{figure}

All modes with wave numbers such that $k v_s/\omega_p<\gamma_s^{-3/2}$ are unstable, and the most unstable mode, i.e., that which grows with the maximum growth rate, is
\begin{equation}
\frac{ k_{\rm f} \hspace{.05cm} c }{ \omega_p } = \sqrt{\frac{3/8}{ (v_s/c)^2 \gamma_s^{3}}}
\quad
\Rightarrow
\quad
\lambda_{\rm f } = 1.6329 \lambda_{\rm min},
\end{equation}
where $\lambda_{\rm min} = 2\pi v_s \gamma_s^{3/2} / \omega_p$ is the minimum unstable wavelength. 
In Figure~\ref{fig:2stream}, we use the cold-plasma dispersion relation of the two-stream instability to plot the growth rate as a function of $k$ for the two-stream instability: red, green, and black curves are the solutions when $\gamma_s=$ $10$, $20$, and $30$, respectively.
As this plot shows, neighboring modes grow at similar rates, and thus the instability has broad spectral support (we quantify the instability width below).
Even if a simulation does not exactly resolve the fastest growing mode, it will nevertheless resolve a nearby mode that grows almost as fast.

To quantify this consideration, let us define a full-width, half-max of this instability in wave number space, with $\Delta k_{1/2}$ as the width in $k$-space, such that all modes grow within a factor of $0.5$ of the maximum growth rate. For the two-stream instability, this is given by:
\begin{equation}
\Delta k_{1/2} =  \frac{1.07561}{ \gamma_s^{3/2}} \frac{ \omega_p}{v_s}.
\end{equation}
Therefore, any periodic box with longitudinal size, $L_{\parallel}$, such that 
\begin{equation}
L_{\parallel} > \frac{2\pi}{\Delta k_{1/2}} =   1.07561 \lambda_{\rm min},
\end{equation}
will necessarily resolve a mode growing with a rate within a factor of $1/2$ of the maximum growth rate.
Therefore, the necessary condition that the box size is larger than the fastest growing wavelength, i.e., $L_{\parallel} > {\lambda}_{\rm f}$, is sufficient to capture the correct linear evolution of such instability in this case.

\subsection{Narrow spectral support: beam-plasma  instabilities}

A more common type of instabilities for astrophysical plasmas are beam-plasma instabilities. They occur when relativistic plasma beams propagate through denser plasmas, which often define the background frame. 
The two-stream and beam-plasma systems differ in key aspects.
The two-stream systems are typically studied in the center of mass frame, where both streams have the same number density and counter-stream with the same speed.
Therefore, the growth rates are characterized only by $\gamma_s$.

On the other hand, beam-plasma systems are typically studied in the frame of the background plasmas and growth rates are characterized by the beam-to-background number density ratio, $\alpha$, and the Lorentz factor of the beam in the frame of the background plasma,$\gamma_b$.
Because $\alpha<1$ in the background frame, there is no other frame of reference where both beam and background have the same number density.

For an ultra-relativistic, very dilute beam streaming through a plasma, it was shown in \citet{Bret2006} that the unstable wave modes lie in a very narrow wave-band within a small angle from the beam direction. The full range of unstable modes was studied in detail in~\citet{Bret2010}.
We focus on the longitudinal modes to show how narrow spectral support naturally emerges in beam-plasma systems.

\begin{figure}
\center
\includegraphics[width=8.6cm]{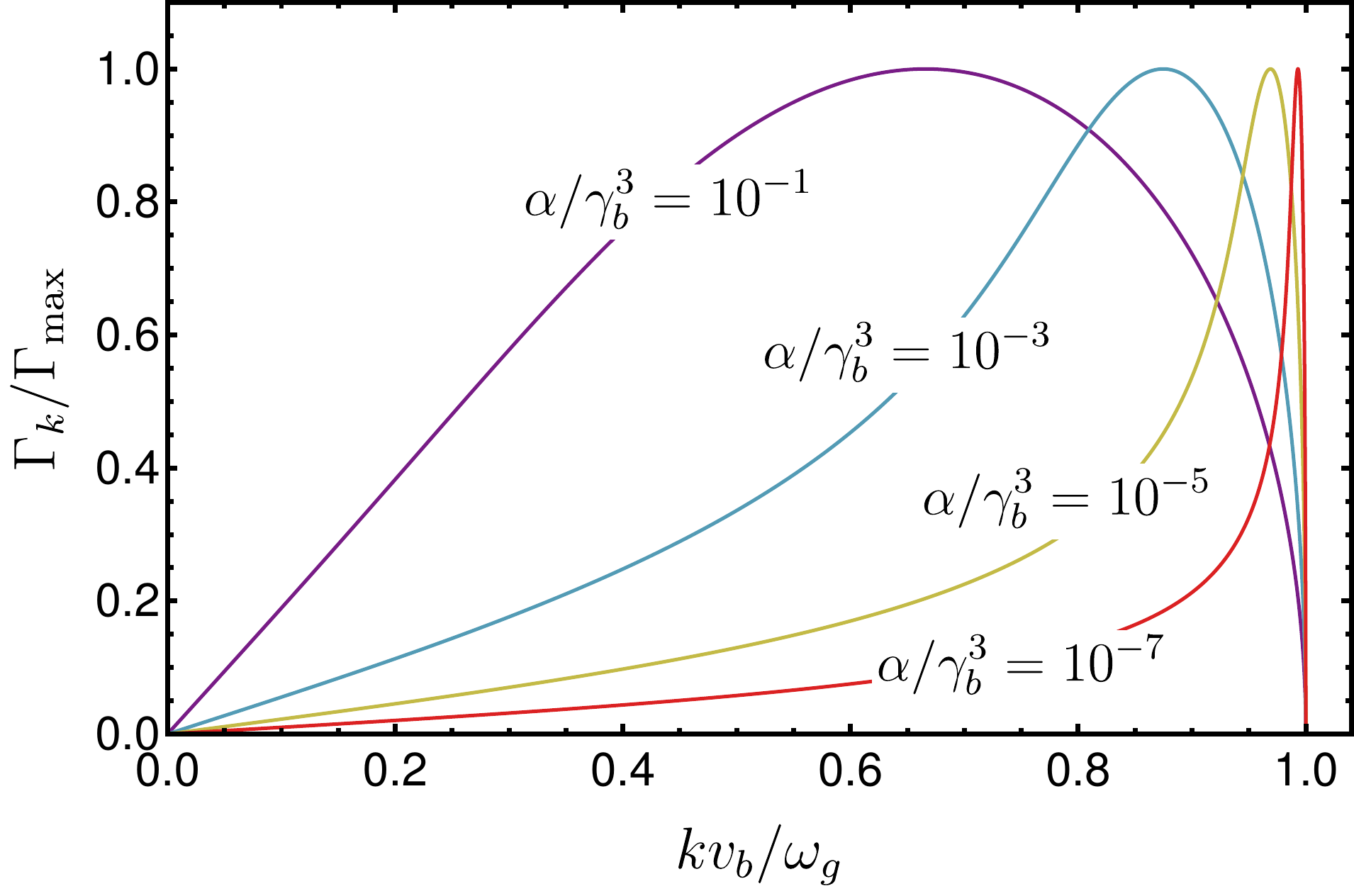}
\caption{
Unstable longitudinal wave modes in beam-plasma systems with different $\alpha/\gamma_b^3$. We show the solutions for the cold limit dispersion relation --  Equation~\eqref{eq:dis_cold}.
\label{fig:Beam-Plasma}
}
\end{figure}

The dispersion relation for longitudinal modes of the beam-plasma system, in the cold limit, is~\citep{fainberg+1969}
\begin{eqnarray}
\left[
 \hat{k}^2 +
\beta_b ^2 \alpha /  \gamma_b +
\beta_b ^2   \left( 1 - \hat{\omega }^2 \right)
\right]^2 
\left[
1-  \frac{1}{\hat{\omega }^2}
- \frac{\alpha /\gamma _b^3 }{  (\hat{\omega }-\hat{k} )^2}
\right]
=
0
\end{eqnarray}
where $\beta_b  = v_b/c$, $v_b$ is the beam velocity, $\gamma_b=(1-\beta_b)^{-1/2}$, $\hat{\omega}=\omega/\omega_g$, $\hat{k} = k_{\parallel}v_b/\omega_g$, $\omega_g $ is the background plasma frequency, i.e., $\omega_g = \sqrt{e^2 n_g/\epsilon_0 m_e}$ and $\alpha=n_b/n_g$, where $n_b$ is the number density of both electrons and positrons beam, and $n_g$ is the number density of background electrons. Background ions are assumed to be immobile and have a uniform density.
The roots of the first part gives only real solutions for $\hat{\omega}$, the stable electromagnetic modes,
while the roots for the second part admit imaginary solutions for $\hat{\omega}$. Thus, the relevant dispersion relation can be written as 
\begin{eqnarray}
1 = \frac{1}{\hat{\omega}^2} + \frac{\alpha/\gamma_b^3}{(\hat{\omega}-\hat{k})^2}.
\label{eq:dis_cold}
\end{eqnarray}
Equation~\eqref{eq:dis_cold} is a quartic polynomial whose discriminant is negative when $\hat{k}^2 < (1+\alpha^{1/3}/\gamma_b)^{3}$. 
For $\hat{k}^2 > (1+\alpha^{1/3}/\gamma_b)^{3} $, the discriminant is positive and all roots are real.
Therefore,
the shortest unstable wavelength is $\lambda_{\rm min} = 2 \pi v_b (1+\alpha^{1/3}/\gamma_b)^{-3/2} /\omega_g$.

The growth rates, i.e., solutions using Equation~\eqref{eq:dis_cold}, are shown in Figure~\ref{fig:Beam-Plasma} for various value of $\alpha/\gamma_b^3$.
For values of $\alpha/\gamma_b^3$ close to unity, the spectral support of unstable wave modes is broad, similar the two-stream system.
For smaller values of $\alpha/\gamma_b^3$, i.e., a more relativistic and dilute beam, the spectral support of the high growth rates narrows considerably.
The spiky nature of the growth rate around the maximally growing mode shown is generic for dilute and relativistic beams.

\begin{figure}
\center
\includegraphics[width=8.6cm]{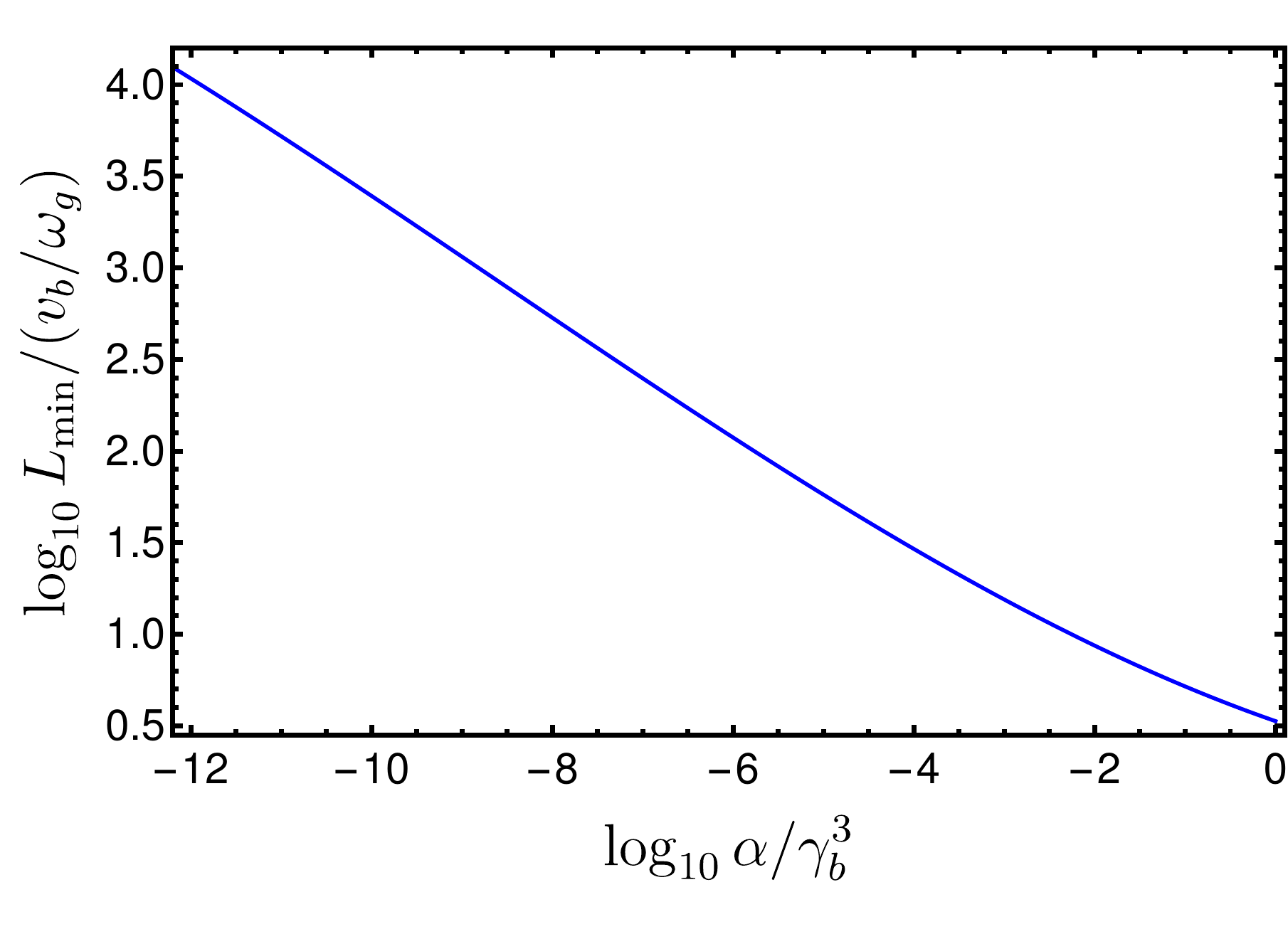}
\caption{
The dependence of the minimum box size, normalized to $v_b/\omega_g$ on $\alpha/\gamma^3_b$. The minimum box size ensures that at least one growing mode with  $\Gamma^{\rm sim}_{\rm max} / \Gamma^{\rm th}_{\rm m} \geq 0.5 $ is resolved.
\label{fig:lmin}}
\end{figure}

The need to resolve such a narrow spectral feature translates into a stringent requirement on the length of the simulation box size.
To derive the requirement on the box size, we again find the full-width half-max, $\Delta k_{1/2}$, of the growth rates, as a function of $\alpha/\gamma^3_b$. 
We then define the minimum longitudinal box-size, $L_{\rm min}=2 \pi / \Delta k_{1/2}$, needed to ensure that at least one wave mode within $\Delta k_{1/2}$ can be resolved. In Figure~\ref{fig:lmin}, we show the dependence of the minimum longitudinal box size on beam parameters ($\alpha/\gamma_b^3$).
Approximately, we find~\footnote{
The quantitative result of Equation~\eqref{eq:lmin} is consistent with the scaling for $\Delta k$ stated in~\citep{fainberg+1969} for the case of $\alpha \ll 1$.
}
\begin{equation}
\label{eq:lmin}
L_{\rm min}
\approx 
\frac{  1.15008  v_b}{\omega_g} \left(\frac{\alpha }{\gamma _b^3}\right)^{-1/3}.
\end{equation}

For a simulation with a box size $L$, such that $\lambda_{\rm min} < L<L_{\rm min}$, it is possible to {\it tune} the box size so that a wave mode within $\Delta k_{1/2}$ is resolved\footnote{When $\lambda_{\rm min} <L<L_{\rm min}$, tuning can allow only one wave mode to grow within $\Delta k_{1/2}$.}.
We explore that possibility below.

\section{Resolving linear beam-plasma instabilities}
\label{sec:simulations}

Here, we explicitly demonstrate the implications of the narrow spectral support on the growth rates found in PIC simulations. 
We find quantitative agreement between the prediction of the maximum growth in simulations and the results from the simulations.

\subsection{Importance of spectral resolution}

\begin{figure} 
\center
\includegraphics[width=8.6cm]{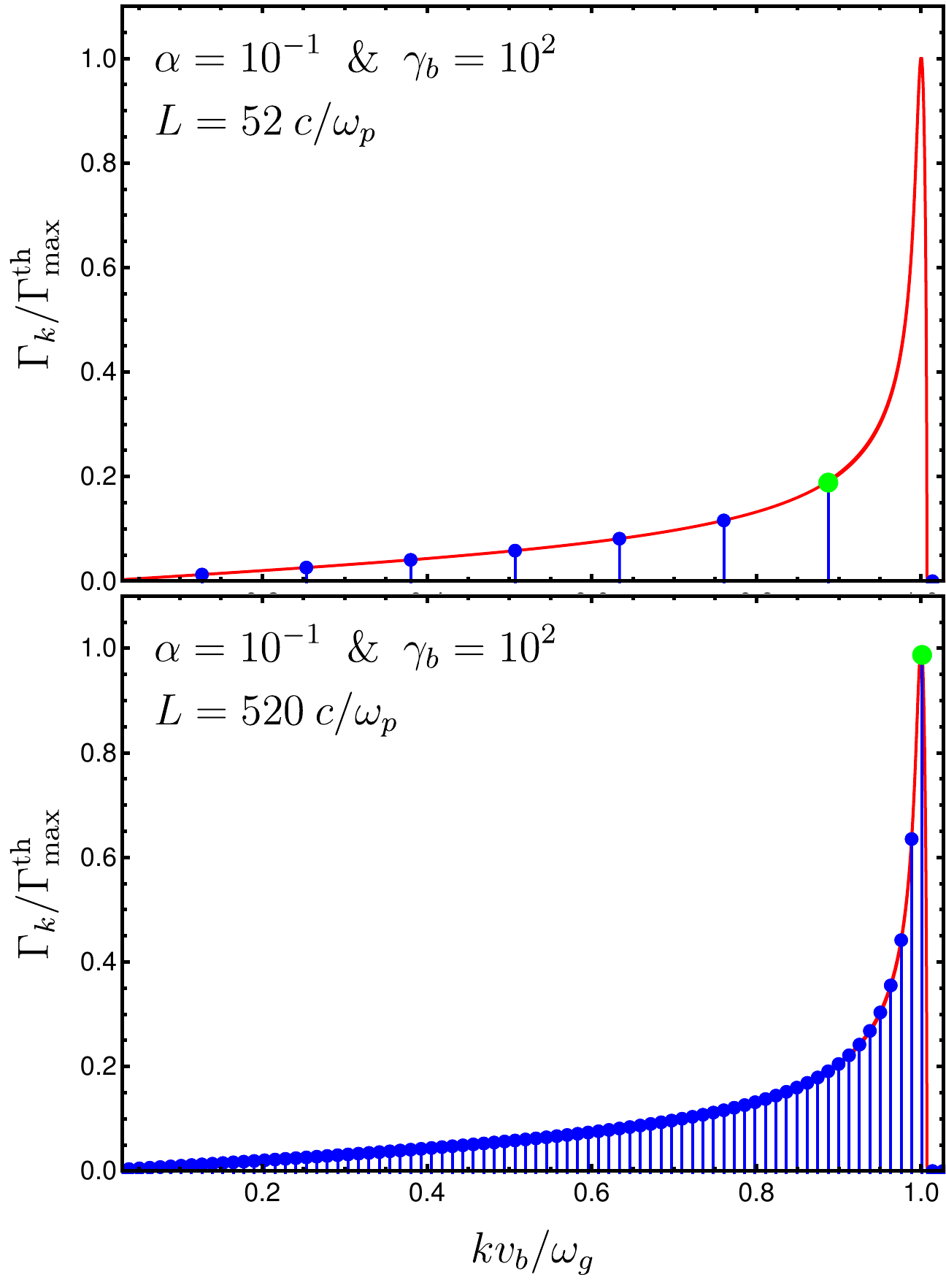}
\caption{
Importance of spectral resolution when simulating the linear phase of cold beam-plasma longitudinal instabilities. The red curves show the normalized growth rate, $\Gamma_k/\Gamma^{\rm th}_{\rm  max}$, of the unstable wave modes when $\alpha/\gamma_b^3=10^{-7}$.
The blue dots represent the growth rates of the wave modes resolved in simulations with box sizes $52$ $c/\omega_p$ (top) and $520$ $c/\omega_p$ (bottom).
The green points indicate the maximum normalized growth rate in simulations with such box sizes, $\Gamma^{\rm sim}_{\rm max}/\Gamma^{\rm th}_{\rm  max}$.
\label{fig:k-res1}}
\end{figure}

The growth rate, $\Gamma_k = \Im{(\hat{\omega_k})}$, of the unstable modes when $\alpha/\gamma^3_b=10^{-7}$ are shown by the red curves in Figure~\ref{fig:k-res1}.
Superimposed are the growth rates of the unstable modes resolved by a simulation box (shown by the points) of lengths $L=52 c/\omega_g$ (top) and $L=520 c/\omega_g$ (bottom).
The maximum growth rate accessible to the simulation, $\Gamma_{\rm max}^{\rm sim}$, is indicated by the green point.
For the smaller box, which spectrally under-resolves the instability, $\Gamma_{\rm max}^{\rm sim}$ is much smaller than that implied by the linear analysis.
In contrast, for the larger box, which contains many modes within the peak of the instability, the two growth rates are similar.

\begin{figure}
\center
\includegraphics[width=8.6cm]{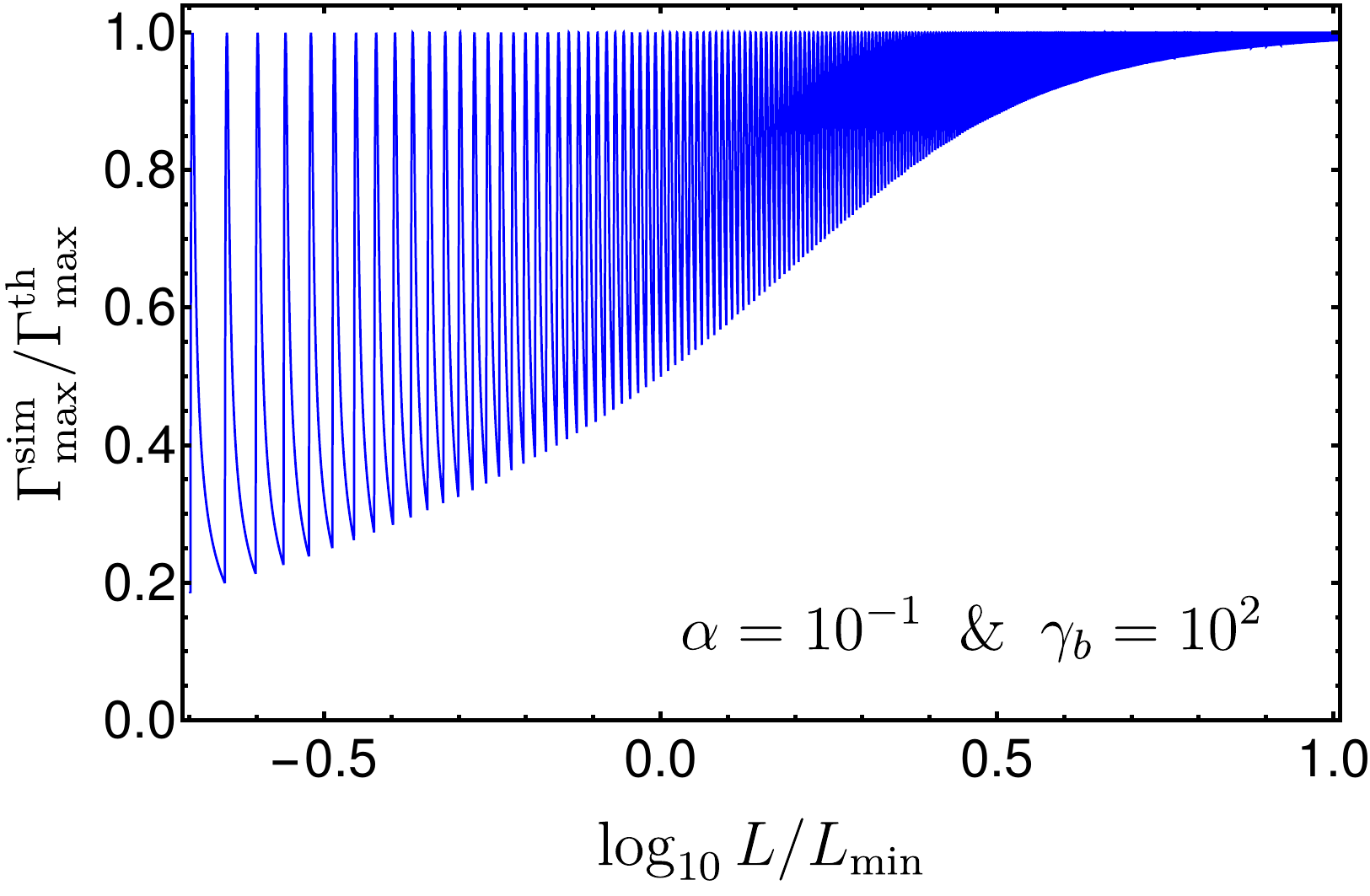}
\caption{
The dependence of the maximum growth rate in simulations on the box size for the case of $\alpha/\gamma_b^3 = 10^{-7}$.
\label{fig:gsim}}
\end{figure}

The ratio of the maximum growth rate accessible in a simulation,  $\Gamma^{\rm sim}_{\rm max}$, to the theoretical maximum growth rate, $\Gamma^{\rm th}_{\rm max}$, is shown as a function of box size for $\alpha/\gamma_b^3=10^{-7}$ in Figure~\ref{fig:gsim}.
At small box sizes, the range of $\Gamma^{\rm sim}_{\rm max}/\Gamma^{\rm th}_{\rm max}$ is large, indicating that the maximum growth rate is frequently missed.
At large box sizes, the range becomes small, as a result of the increased spectral resolution, i.e., the instability is resolved.
Note that, even at low resolution, it is possible for the fastest-growing mode to land at maximum theoretical growth rate, e.g., as shown in Figure \ref{fig:k-res2}.
As mentioned above, this shows how we might attempt to {\it tune} simulations to resolve the fastest growing modes by design; we explore this in the following section.

\subsection{Numerical verification}

\begin{figure} 
\center
\includegraphics[width=8.6cm]{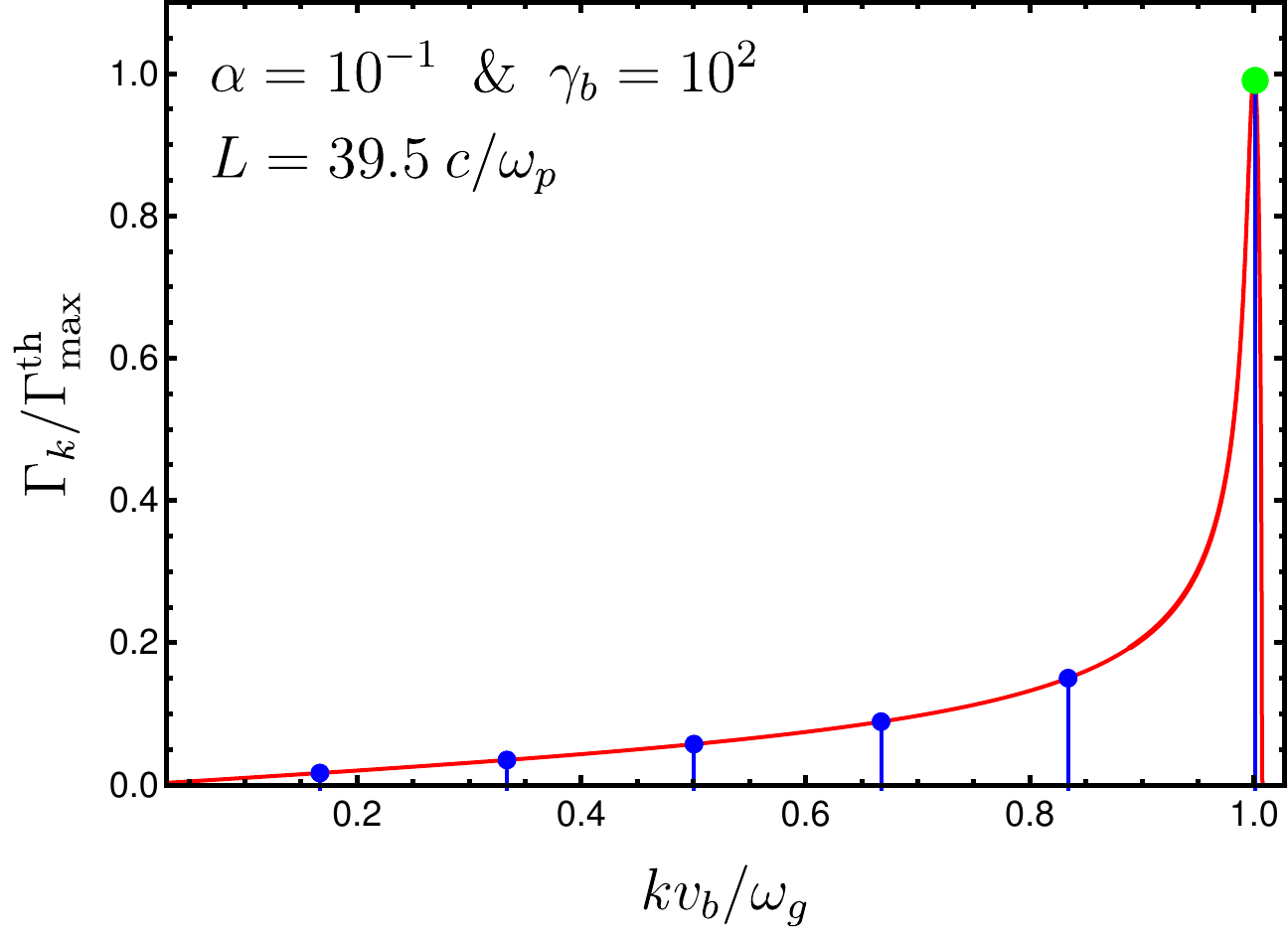}
\caption{
Tuning spectral resolution in simulation: the normalized growth rates $\Gamma_k/\Gamma^{\rm th}_{\rm  max}$, of the unstable wave modes when $\alpha/\gamma_b^3=10^{-7}$ (red curve), the blue dots are the growth rates of the wave modes resolved by a simulation with box size $L = 39.5$ $c/\omega_p$.
Green point indicates the maximum normalized growth rate in the simulation with such box size, $\Gamma^{\rm sim}_{\rm max}/\Gamma^{\rm th}_{\rm  max}$. 
\label{fig:k-res2}
}
\end{figure}

Here, we use ab initio simulations to demonstrated the importance of the spectral resolution. 
We run a series of simulations where we progressively increase spectral resolution, which we list in Table~\ref{table:sim-para}, and a simulation with a small, but tuned, box size (final row of Table~\ref{table:sim-para}) such that a mode growing at a rate very close to the maximum growth rate is resolved in our periodic box, as shown in Figure~\ref{fig:k-res2}.

\begin{deluxetable}{
cccc
}
\tablewidth{8.6cm}
\tabletypesize{\footnotesize}
\tablecolumns{4} 
\tablecaption{
Electrostatic Beam-plasma instability simulations for $\alpha = 0.1$ and $\gamma_b=100$.
\label{table:sim-para}
}
\tablehead{
$L_c$  \tablenotemark{a}					& 
$L/L_{0}$\tablenotemark{b}					& 
$\Gamma^{\rm sim}_{\rm max} / \Gamma^{\rm th}_{\rm max} $\tablenotemark{c} &
$k^{\rm sim}_{\rm max} c / \omega_p$  \tablenotemark{d}
}
\startdata
\rule{0pt}{8pt}
$026$ & $0.1$ 	& $ 0.1162 $ & $0.724983$ 
\\
\rule{0pt}{8pt}
$052$ & $0.2$ 	& $ 0.1908 $ & $0.845813$
\\
\rule{0pt}{8pt}
$130$ & $0.5$ 	& $ 0.3550 $ & $0.918312$
\\
\rule{0pt}{8pt}
$260$ & $1.0$ 	& $ 0.6351 $ & $0.942478$
\\
\rule{0pt}{8pt}
$520$ & $2.0$ 	& $ 0.9897 $ & $0.954561$
\\
\rule{0pt}{8pt}
$1040$ & $4.0$ 	& $ 0.9897 $ & $0.954561$
\\
\rule{0pt}{8pt}
$39.5$ &$0.152$ & $ 0.9926 $ & $0.954408$
\enddata
\tablenotetext{a}{The box size, $L$, in units of skin depth, i.e., $ L_c = L ~ \omega_p / c$, where, $\omega_p$ is the plasma frequency associated with all plasma particles: beam and background particles.}
\tablenotetext{b}{$L_0 = 260 \text{ }c/\omega_p \sim L_{\rm min}(\alpha/\gamma_b^3=10^{-7}) $.}
\tablenotetext{c}{The maximum growth rate predicted for simulations, due to different spectral resolutions $\Gamma^{\rm sim}_{\rm max} $, normalized to the maximum growth rate, $\Gamma^{\rm th}_{\rm max} 
= 3.036 \times 10^{-3} \text{ } \omega_p$, found by solving the dispersion relation in Equation~\eqref{eq:dis_cold}.}
\tablenotetext{d}{The normalized fastest growing wave mode in simulations.
}
\end{deluxetable}

For all simulations here, we use SHARP-1D~\citep{sharp} with fifth order interpolation, $W^5$, to improve the conservation of energy in simulations while conserving the total momentum exactly.
Using SHARP with $W^5$ is essential to avoid the excessive numerical heating typical in most available PIC codes \citep[see][for illustration]{sharp}.
Predictions of the maximum growth in different simulations are given in Table~\ref{table:sim-para} (green points in Figures~\ref{fig:k-res1} and~\ref{fig:k-res2}).
Here, we show a quantitative agreement between simulations' results and these predictions for the case of $\alpha = 0.1$ and $\gamma_b=100$.

\begin{figure*}
\center
\includegraphics[width=18cm]{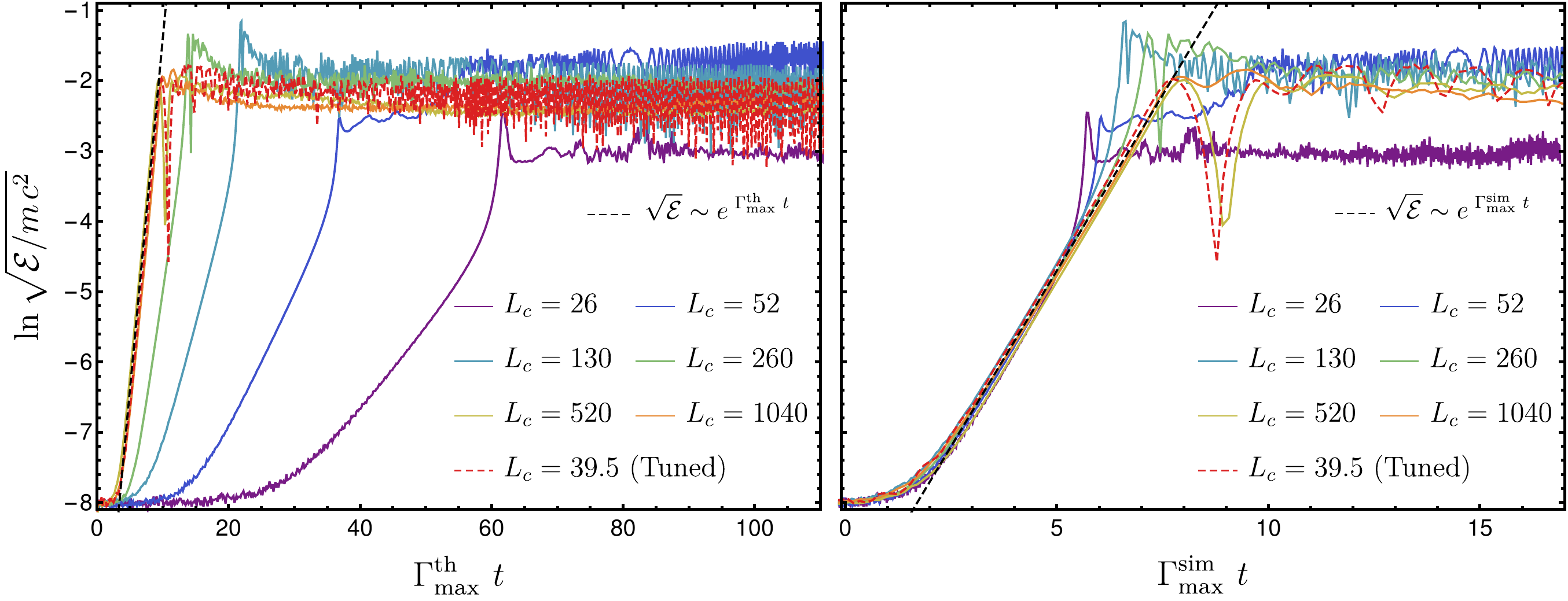}
\caption{
Simulation results: the growth of the total potential energy per computational particle, $\mathcal{E}$, normalized to the rest mass energy of a computational particle,  $m \hspace{.05cm} c^2$. 
Left: time is normalized with the maximum growth rate, i.e., the solution of Equation~\eqref{eq:dis_cold}.
Right: time is normalized with the maximum growth rate predicted in a simulation as shown in Figures~\ref{fig:k-res1} and \ref{fig:k-res2}. Here, $L_c  = L ~ \omega_p/c$.
Because the growth in all simulations starts from the Poisson noise, the times are shifted in different simulations to allow for comparisons.
\label{fig:PE}}
\end{figure*}

\begin{figure*}
\center
\includegraphics[width=18cm]{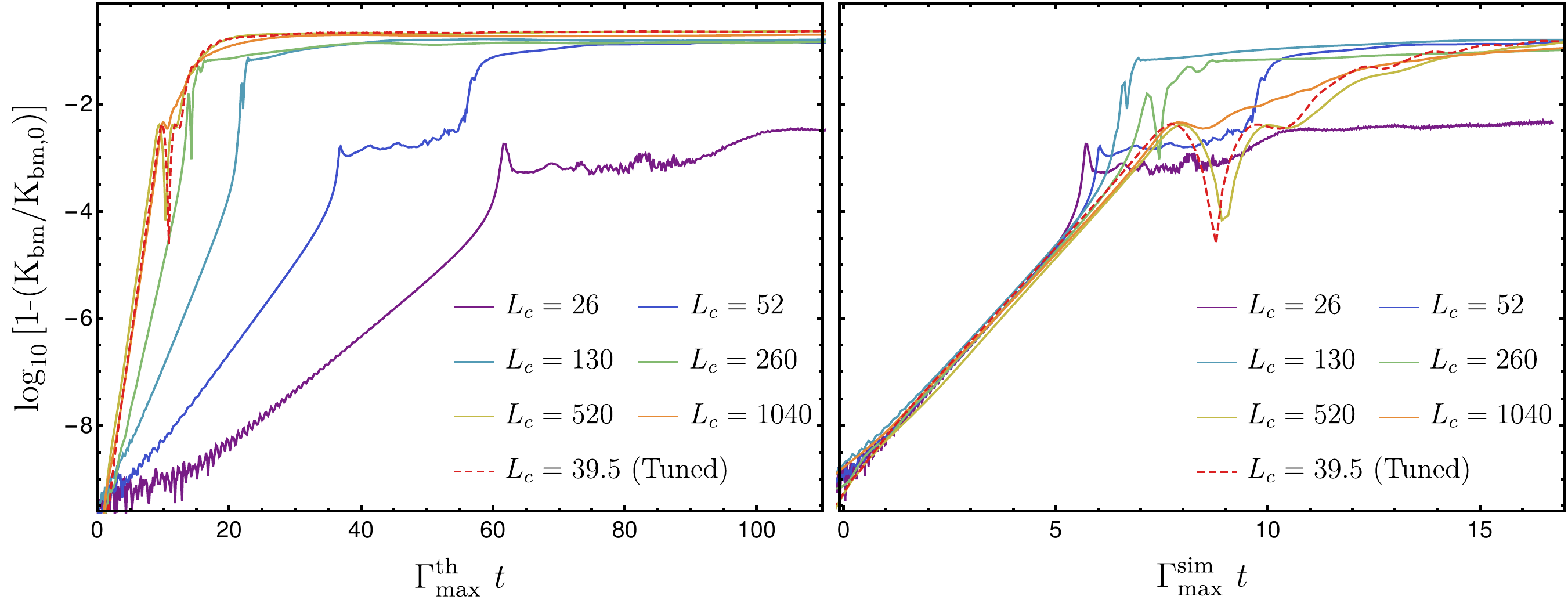}
\caption{
Simulation results: the evolution of the kinetic energy loss of the beam.
Left: time is normalized with the maximum growth rate, i.e., the solution of Equation~\eqref{eq:dis_cold}.
Right: time is normalized with the maximum growth rate predicted in a simulation as shown in Figures~\ref{fig:k-res1} and \ref{fig:k-res2}.
Here, $L_c  = L ~ \omega_p/c$.
Because the growth in all simulations starts from the Poisson noise, the times are shifted in different simulations to allow for comparisons.
\label{fig:KE}}
\end{figure*}

We fix both spatial and momentum resolutions in all simulations: $\Delta x = 0.05 c / \omega_p$ and $N_{\rm pc} = 1650$, where $\Delta x$ is the grid cell size and $N_{\rm pc}$ is the number of computational particles per cell.\footnote{Note, the total plasma frequency, $\omega_p$, is related to the background plasma frequency, $\omega_g$, as, $\omega_p / \omega_g = \sqrt{1+\alpha}$.
}
We start all simulations with a uniform distribution of both electron and positron beam particles, propagating with $\gamma_b=100$ (in the same direction) through an initially uniform background of electrons and a fixed background of ions. The initial normalized temperature of the background electrons is $\theta_g = k_B T_g/m \hspace{.05cm} c^2 = 10^{-3}$, and both types of beam particles have an initial co-moving temperature of $\theta_b = k_B T_b/m \hspace{.05cm} c^2 = 4\times10^{-3}$, where, $m \hspace{.05cm} c^2$ is the rest mass energy of a computational particle.
The rest of the simulation parameters for the  different simulations are given in Table~\ref{table:sim-para}.

Using the solution of the cold-limit dispersion relation, Equation~\eqref{eq:dis_cold}, we find that the fastest growing mode is $k^{\rm th}_{\rm m} = 0.95353  c/\omega_p$ and its exponential growth rate is $\Gamma^{\rm th}_{\rm max} = 0.0030364 \text{ } \omega_p$.
Because this mode is not typically resolved in a given simulation, simulations with different spectral resolutions will have different maximally growing wave modes $k^{\rm sim}_{\rm m}$, and thus, a non-unity $\Gamma^{\rm sim}_{\rm max}/\Gamma^{\rm th}_{\rm max}$ 
(listed in Table~\ref{table:sim-para}).

\begin{figure}
\center
\includegraphics[width=8.6cm]{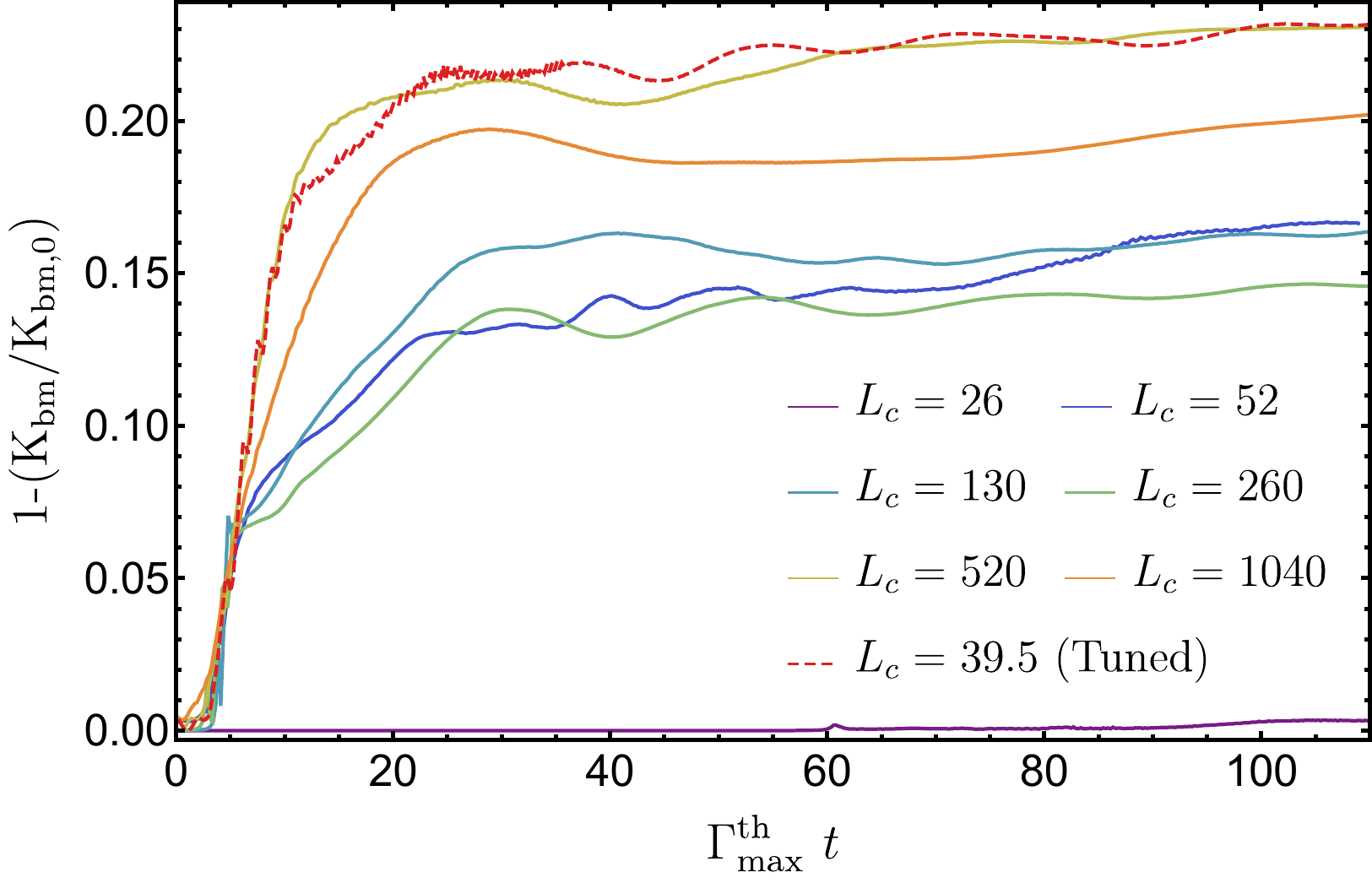}
\caption{
{ 
The evolution of the beam energy loss on a linear scale to better show the impact on the nonlinear saturation.
The times are shifted in different simulations to allow for comparisons.
The saturated energy transfer varies over three orders of magnitude and is not monotonically related to the box size (i.e., degree of under-resolution).
}
\label{fig:KE-linear}}
\end{figure}

In Figures~\ref{fig:PE} and \ref{fig:KE}, we show the evolution of the electric potential energy and beam energy loss respectively.
In the left panels, time is normalized with the maximum growth rate, i.e., the solution of Equation~\eqref{eq:dis_cold}.
In the right panels, time is normalized with the maximum growth rate predicted in simulations. 
Because the growth in all simulations starts from the Poisson noise, the times are shifted in different simulations such that the linear growth are aligned.
These are shown on a linear scale in Figure \ref{fig:KE-linear} to better show the impact on the saturation of different box sizes.

It is difficult to make general statements regarding the impact of insufficient spectral resolution on the nonlinear saturation of the beam plasma system.
However, it is clear from the late-time evolution in Figures~\ref{fig:PE}--\ref{fig:KE-linear}, that both the saturation of the mode amplitudes and the energy transfer to the background can be severely reduced if the instability is badly under-resolved.
We would anticipate that this would be even more severe in cases where the beams are also energetically sub-dominant, and thus incapable of heating the background to relativistic temperatures.
Clearly visible in Figure \ref{fig:KE-linear} is the wide variation in the nonlinear saturation of the instabilities; we did not identify any general relationship between the error in the saturated energy transfer with the degree of under-resolution\footnote{\
We note that larger boxes at fixed $N_{\rm pc}$ and $\Delta x$ lead to larger levels of Poisson noise~\citep{sharp}, and when $N_{\rm pc}$ is increased from 1650 to 16500 to decrease the Poisson noise, the level of nonlinear saturation of $L_c = 1040$ simulation increases from $\sim 20$\% to $\sim 23$\%.}.

Both tuning and using large box sizes result in similar evolutions in the linear regime.
The reasons for this are, first, the longitudinal unstable  modes are insensitive to the background heating resulting from the energy exchange due to these instabilities, i.e., despite the physical heating, the simulation box will still be able to resolve a wave mode growing with at a rate comparable to the $\Gamma_{\rm max}^{\rm th}$. 
Second, the very narrow spectral support shown in Figure~\ref{fig:k-res1} means that the linear-regime evolution of the system will largely be dictated by the fastest-growing wave mode.

\section{Concluding remarks}
\label{sec:conclusion}

A large subset of astrophysically important plasma instabilities, notably the beam plasma instabilities, have narrow spectral support, i.e., only a small subset of linearly unstable wave numbers grow at rates comparable to the maximum growth rate.
This is especially true for the unstable longitudinal components  of relativistic, dilute beam-plasma systems.

Accurately resolving these in numerical simulations, even in the linear regime, places stringent requirements on the spectral resolutions of the simulations.
This, in turn, places {\it lower} limits on the size of the simulation box.
Typically, for dilute, relativistic plasma beams, this requires simulation boxes much larger than the typical unstable mode wavelengths.
That is, requiring that the simulation has sufficient spectral resolution to resolve the full-width, half-max of the beam-plasma instability necessitates box sizes many orders of magnitude larger than the wavelength of the fastest-growing modes.

Alternatively, smaller simulation boxes can be used, but they would need to be tuned. We note that, while this was possible in 1D, it is unlikely to be the case in 2D and 3D.
The impact of beam temperature on the maximally growing mode was studied by \citet{Bret-2010-tp} for longitudinal, transverse, and oblique instabilities.
For the relativistic, dilute case, the fastest growing longitudinal mode is surprisingly insensitive to the temperature (see Figure~16 of \citealt{Bret-2010-tp}); this explains the success of our tuned simulation.
However, the fastest growing transverse and oblique modes are much more sensitive to temperature.
Therefore, even moderate heating can lead to significant evolution in the wave number of the fastest-growing mode, precluding tuning.

Using explicit PIC simulations, we have demonstrated that insufficient spectral resolution fully explains the reduced linear growth rates in 1D beam-plasma systems.
That is, it is possible to quantitatively reproduce the simulated growth rates with those anticipated by the limited compliment of oscillation modes present in a simulation associated with fixed box size.
These compare well with results of similar simulations in the literature \citep[see, e.g.,][]{gremillet2007,Bret2010,lorenzo2014,kempf2016, Rafighi2017}, where the maximum growth rates are found to be smaller than anticipated by the linear theory, in some cases by up to an order of magnitude.
This suggests that the reduced linear growth rates are a consequence of insufficient spectral resolution, i.e., simulation boxes that are too small.
   
While an extensive study of the impact of insufficient spectral resolution on the nonlinear saturation of the beam-plasma system has yet to be performed, it is clear that it does have an effect.
The lowest-resolution simulation that we ran, under-predicting the linear growth rate by nearly an order of magnitude, transferred two orders of magnitude less energy than the simulations that did spectrally resolve the instability.
This raises concerns about the robustness of the conclusions that may be drawn from published simulations of relativistic beam-plasma systems.

For multidimensional (2D or 3D) simulations, in the cold limit, the width of the  instability of the oblique modes is typically larger than that for the longitudinal modes.
However, when finite temperature effects are included, the growth maps of the instabilities become much narrower~\citep[see, e.g.,  the various panels in Figure 17 of ][]{Bret2010}.
Hence, the spectral resolution requirements imposed on box size are likely to be exacerbated in the warm cases.

\setcitestyle{notesep={; }}
Another related issue when considering the multidimensional simulations is that resolving only the oblique modes is know to be insufficient. 
In practice, all unstable modes grow simultaneously and impact the subsequent nonlinear evolution. 
This has also been shown in~\citet[][see especially the discussion in Section V.D]{Bret2010}.
Therefore, if the longitudinal instability imposes a stronger restriction on the simulation setup, it must continue to be respected in higher dimensions as well.
That is, our 1D analysis is essential to ensure that the linear evolution and the nonlinear saturation are correctly captured in higher-dimensional simulations as well.
\setcitestyle{notesep={, }}
 
\section*{Acknowledgments}

M.S. and A.E.B. receive financial support from the Perimeter Institute for Theoretical Physics and the Natural Sciences and Engineering Research Council of Canada through a Discovery Grant. Research at Perimeter Institute is supported by the Government of Canada through Industry Canada and by the Province of Ontario through the Ministry of Research and Innovation.
P.C. gratefully acknowledges support from the NASA ATP program through NASA grant NNX13AH43G, and the NSF through grant AST-1255469.
C.P. gratefully acknowledges support by the European Research Council through ERC-CoG grant CRAGSMAN-646955 and by the Klaus Tschira Foundation.
E.P. gratefully acknowledges support by the Kavli Foundation.

Support for AL was provided by an Alfred P. Sloan Research Fellowship, NASA ATP Grant NNX14AH35G, and NSF Collaborative Research Grant \#1411920 and CAREER grant \#1455342.

\bibliography{refs}
\bibliographystyle{apj}

\end{document}